\documentclass[aps,twocolumn,prc,superscriptaddress,noshowpacs,nofootinbib,noshowkeys,floatfix]{revtex4-1}
\usepackage[colorlinks=true,linktocpage=true,linkcolor=blue,citecolor=blue]{hyperref}
\usepackage[usenames,dvipsnames]{color}
\usepackage{amsmath, amssymb}
\usepackage{multirow}
\usepackage{longtable}
\usepackage{comment}
\usepackage{color}
\usepackage{xcolor}
\usepackage{xspace}
\usepackage{cleveref}
\usepackage[normalem]{ulem}  % \sout{old text} for strikeout
\usepackage{graphicx,lipsum}
\usepackage{ragged2e}
\newcommand{\bef}{\begin{figure}}
\newcommand{\eef}{\end{figure}}
\newcommand{\bc}{\begin{center}}
\newcommand{\ec}{\end{center}}

\newcommand{\be}{\begin{equation}}
\newcommand{\ee}{\end{equation}}
\newcommand{\bea}{\begin{eqnarray}}
\newcommand{\eea}{\end{eqnarray}}

\begin{document}

\title{{Centrality Dependence of the Balance Functions for Identified Particles in Pb--Pb Collisions Using Pythia + Angantyr}}

\author{Rashi Gupta}
\email{phd2101151026@iiti.ac.in}
\affiliation{Department of Physics, Indian Institute of Technology Indore, Simrol, Indore 453552, India}
\author{Ankhi Roy}
\email{ankhi@iiti.ac.in}
\affiliation{Department of Physics, Indian Institute of Technology Indore, Simrol, Indore 453552, India}

\begin{abstract}
In this paper, we study the balance functions for pions, kaons, protons in Pb-Pb collisions at \(\sqrt{s_{NN}}\) = $2.76 $  TeV using the Pythia8.3 + Angantyr model.  The balance functions is evaluated through two-particle azimuthal angular correlations  (\(\Delta \varphi, \Delta \eta\)) between particle and antiparticle.  Correlations are constructed for  \(\pi \pi\), \(K K\), \(p p\), and their dependence on collision centrality is investigated. The results indicate that the balance function for pions is narrower compared to kaons and protons. Notably, the pion balance functions  width decreases from peripheral to central collisions, while the balance functions widths for kaons and protons remain nearly
unchanged. For the Monash 2013 tune used in this study, Pythia8.3 + Angantyr describes peripheral collisions reasonably well but does not quantitatively reproduce central Pb–Pb data. This suggests that an improved description of central Pb–Pb collisions may require a dedicated heavy-ion tuning of the Angantyr framework. We further explore the influence of resonance decays and collective  effects by incorporating multi-parton interactions and color reconnection into the analysis. Owing to resonance effects and Bose–Einstein correlations, a dip at $\Delta \eta $ = 0 and $\Delta \varphi $ = 0 is observed for pions and kaons.

\end{abstract}
\date{\today}
\maketitle

\section{Introduction}
 In hadron–hadron and heavy-ion collisions, a large variety of particles are produced, and their production is governed by several conservation laws, such as electric charge (Q), baryon number (B), strangeness (S), beauty (b), and charm (c). In strong interactions, all these quantum numbers are conserved, whereas in weak decays, strangeness and quark-flavor quantum numbers are not conserved. Since particle production in hadron–hadron and heavy-ion collisions proceeds predominantly via strong interactions, these quantum numbers are conserved at the production stage. The balance functions (BF)~\cite{Bialas2004, STAR_BalanceFunction, STAR_Balance_Adamczyk, STAR_Balance_Aggarwal, ALICE_Balance_Adam, ALICE_Balance_Abelev, chargebalance1, electricbalance, SystemsizeBalance} is a tool that measures charge–anticharge correlations and provides information about the production time of quark–antiquark pairs. The BF distinguishes between different stage particle production: if the correlated charge--anticharge pairs are produced in the early stage of the collision, the balance functions becomes broader due to diffusion and collective expansion, whereas production in the later stage leads to a narrower BF \cite{Pratt2002, narrowbalance}. This behavior is illustrated in figure 1.  Pions, composed of light quarks, are typically produced at later stages and therefore exhibit a narrower balance functions, whereas kaons, containing a strange quark, are produced earlier and display a comparatively broader balance functions. In heavy-ion collisions, a quark–gluon plasma (QGP) is formed~\cite{HeavyIon,QGP, STARQGP, Qgp3}. If charged particles are produced at an early stage, they carry information about the medium; hence, the balance functions is used to measure the QGP lifetime and particle formation time~\cite{Bass2000}. In addition to production time, the BF is also influenced by the strength of radial flow~\cite{radialflow1, balanceflow, Radialcorrelation}. Radial flow imparts a collective outward momentum boost to particles~\cite{Radial, collective, collectiovehydro}. A stronger radial flow results in tighter correlations between balancing charges, producing a narrower BF. Thus, the BF width depends not only on the quark production time but also on the radial flow. For instance, in central heavy-ion collisions, a large number of nucleons participate in the interaction, while in peripheral collisions, only a few nucleons are involved. Consequently, the partonic stage takes longer to evolve into the hadronization stage in central collisions compared to peripheral collisions ~\cite{Balancefunction,Balance}. The stronger radial flow developed in central collisions further contributes to the narrowing of the BF. Therefore, when moving from central to peripheral collisions, the BF becomes broader due to both the shorter evolution time and the reduced radial flow. \\
Balance functions measures the  two-particle azimuthal angular correlations (\(\Delta \varphi, \Delta \eta\)) between the  like-sign (LS) and unlike-sign (US) pair~\cite{correlation, centerlitycorrelation, ALICE2024, LHCb2023,Airapetian2015, Bozek2012}. US and LS correlations represent the correlations between particles with opposite charges and same charges, respectively. Here, $\Delta\varphi$ denotes the difference in azimuthal angle, while $\Delta\eta$ represents the difference in pseudo-rapidity 
between the two particles. 
In this article we study the balance function for \(\pi \pi\), \(K K\), \(p p\) using Pythia8.3 + Angyantyr event generator in Pb-Pb  collisions at \(\sqrt{s_{NN}}\) = $2.76$ TeV~\cite{Agantyr1,particleproduction, centerlityATLAS}. We investigate the effect of resonance decays~\cite{Boyle2024, balanceresonance3} on pion and kaon balance functions and also study the impact of Bose--Einstein correlations~\cite{Bose} (BEC), which influence the BF of both $\pi\pi$ and $KK$ pairs. 
These correlations arise for like-sign identical ~\cite{Identicle} bosons emitted from the same source.  \\
In the next section, we describe the event generation and analysis methodology. Then, Section III presents the results and discussion, and finally, Section IV presents the summary and conclusions
\begin{figure}[h!]
  
     \includegraphics[width=0.53 \textwidth]{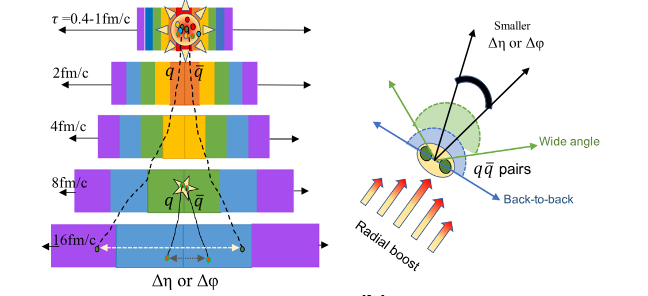} 
  
     \caption{Charge--anticharge pair production at different times ($1$~fm/$c$ and $8$~fm/$c$). Early production results in a broader $\Delta\eta$ and $\Delta\varphi$ distribution, whereas late production leads to a narrower $\Delta\eta$ and $\Delta\varphi$ distribution. The effect of radial flow on narrowing the balance func-
tion is shown in the right-side panel.}

    \label{fig:enter-label}
\end{figure}

\section{Event generation and methodology}
\label{section2}

The Pythia8 event generator is used to study proton–proton and proton–lepton collisions. Since the Pythia\cite{Pythia2, Pythia3, Pythia1, Pythia4, monash13, Campbell:2022qmc} event generator cannot study heavy ion collisions, the Pythia8 + Angantyr~\cite{Angantyr, Agantyr1, eventgenerater} event generator is used, which does not include a medium description. In the absence of a dense medium, Pythia8+Angantyr creates a partonic state  that hadronizes to produce the particles in the final state. Here we used 8.309 version of Pythia8+Angantyr. It is used to study heavy ion collisions like  p-Pb and Pb-Pb. It includes soft and hard scattering processes and  multiparton  interaction (MPI)~\cite{MPI, MPI1, MPI2, MPI3,MPI4}, initial state radiation (ISR), final state radiation (FSR), color reconnection (CR)~\cite{Gieseke:2012ft, Bierlich:2015rha} and resonance effect~\cite{ALICEresonance}. For low momentum transfer, soft scattering is used, while for high momentum transfer, hard scattering.  Hadronization in Pythia is based on Lund string fragmentation model\cite{String, String1} for parton fragmentation.  \\
We  used 300 million event for Pb-Pb at \(\sqrt{s_{NN}}\) = $2.76$ TeV. This analysis  SoftQCD:all= on was utilised for the pion, kaon, and proton. For the  study of collective effect we apply with MPI and with CR. The overall trends and conclusions remain unchanged. Pions and kaons can originate from both resonant and non-resonant sources. In this study, we focus on the resonance contributions \cite{reso}, specifically from the decays of the $\rho^0$, and $\omega$ mesons for pions, and the $\phi$ meson for kaons. The relevant decay channels considered are:
\begin{align}
\rho^0 &\rightarrow \pi^+ + \pi^- \\
\omega &\rightarrow \pi^+ + \pi^- \\
\phi &\rightarrow K^+ + K^-
\end{align}
Pythia8.3+Angantyr does not include BEC in its default setting. To study the BEC effect in Pythia, we need to enable it using HadronLevel:BoseEinstein = on. In Pb–Pb collisions, the system size is very large. For central collisions, Pythia does not accurately describe the Bose-Einstein correlation because, in central Pb–Pb collisions, the system size is approximately 5 fm~\cite{Systemsize, Systemsize1} for both kaons and pions. However, in Pythia, the system size  cannot be increased beyond 2 fm due to the model limitations. In peripheral collisions, where the system size is smaller, Pythia can describe the effect more reasonably. \\
The balance functions is defined as: 
\begin{align}
        B(\Delta \eta, \Delta \varphi) = \frac{1}{2}(B_{UL}(\Delta \eta, \Delta \varphi) - B_{LS}(\Delta \eta, \Delta \varphi))
  \end{align}
where \( B_{UL} \) denotes the Unlike-sign  balance functions and  \( B_{LS} \) denotes the Like-sign  balance functions and these balance functions is defined as:    \begin{equation}
           B_{UL}(\Delta \eta, \Delta \varphi) = C_{+ -}(\Delta \eta, \Delta \varphi) + C_{- +}(\Delta \eta, \Delta \varphi)
    \end{equation}
      \begin{equation}
           B_{LS}(\Delta \eta, \Delta \varphi) = C_{+ +}(\Delta \eta, \Delta \varphi) + C_{- -}(\Delta \eta, \Delta \varphi)
      \end{equation}

 where \(C (\Delta \eta, \Delta \varphi)\) is defined as:
 \begin{equation}
     C(\Delta\eta, \Delta\varphi)  = \beta \times \frac{1/N_{trigger}^{same}S_{SE}(\Delta\eta, \Delta\varphi)} {1/N_{trigger}^{mix}B_{ME}(\Delta\eta, \Delta\varphi)}
 \end{equation}
           Where  $\beta $  is the normalization yield  for mixed events. The signal distribution is the particle yield of pairs in the same event, given by \\   
            \begin{equation*}
                S_{SE}(\Delta \eta ,  \Delta \phi) =  \frac{d^{2}N^{\rm{same}}}{d\Delta\eta d\Delta\phi} 
            \end{equation*}
                
         where, $ N^{same}$  is the number of pairs within a ( $\Delta\eta$, $\Delta\phi$) bin. The background distribution from mixed-event is given by \\
       \begin{equation*}
              B_{ME}(\Delta \eta , \Delta \phi) =  \frac{d^{2}N^{\rm{mix}}}{d\Delta\eta d\Delta\phi}
       \end{equation*}
    where, $N^{mix}$  is the number of mixed-event pairs.  \\
    The centrality distribution of charged particles for Pb–Pb collisions is shown in figure 2 for different Pythia8.3 tunes using the Angantyr model, in comparison with ALICE data \cite{Balance} for Pb-Pb collision at $\sqrt{s_{NN}} = $ 2.76 TeV . The plots include simulations with MPI and CR, as well as without CR, to evaluate the impact of these effects. As shown in figure 2, turning off CR \cite{CR, CR1} leads to increased particle production. This is attributed to the reduction in string lengths when CR is disabled. The centrality classes used in this analysis, listed in Table~\ref{tab:centrality_CR_comparison}, allow for a detailed study of the multiplicity dependence of particle production.  \\
\begin{figure}[h!]
     \hspace*{-1.3cm} % Shift image 1cm to the left
        \includegraphics[width=1.2\linewidth]{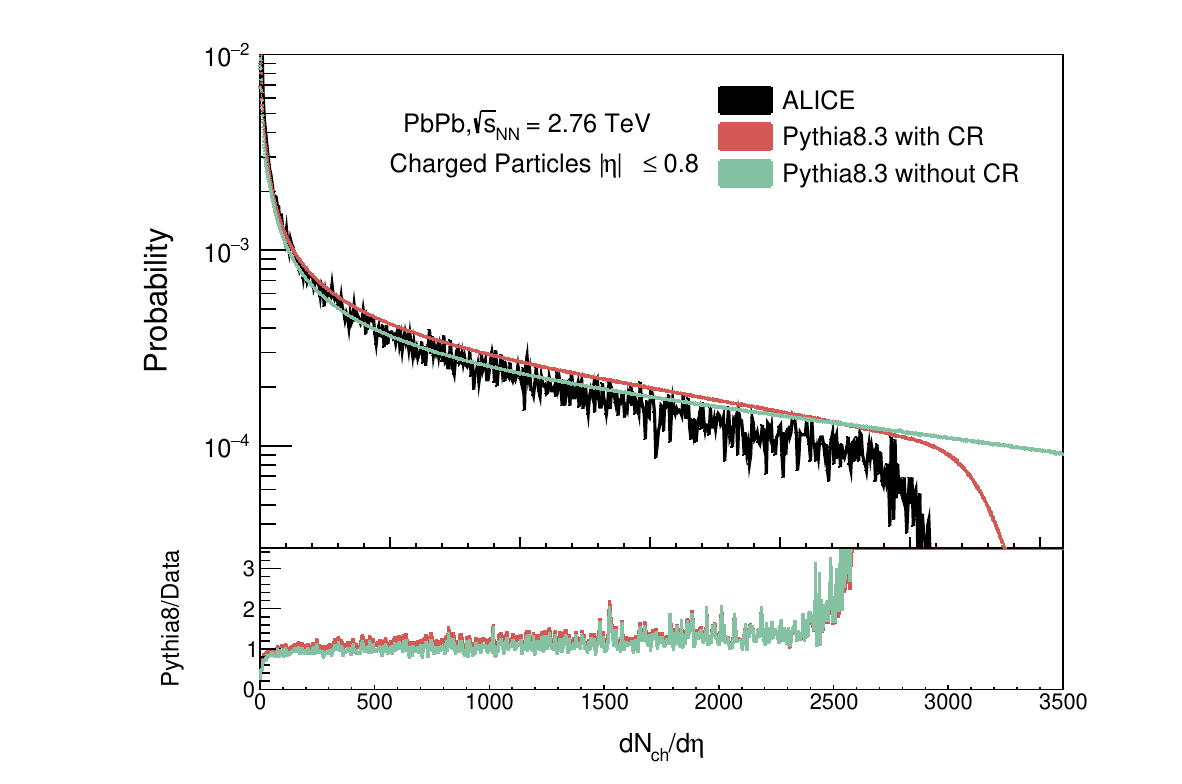}
    \caption{Centrality distribution of charged particles in Pb–Pb collisions from Pythia8.3 + Angantyr using the Monash 2013 tune, compared with data from the ALICE Collaboration, with and without color reconnection (CR).}
    \label{fig:enter-label}
\end{figure}
\begin{table}[htbp]

\caption{Charged particle multiplicities (N\textsubscript{ch}) corresponding to different centrality classes in Pb–Pb collisions are shown for scenarios with and without color reconnection (CR).}
\vspace{0.5em} 
\label{tab:centrality_CR_comparison}
\centering
\resizebox{0.5\textwidth}{!}{%
\begin{tabular}{|c|c|c|c|}
\hline
\textbf{S.No.} & \textbf{Centrality (\%)} & \textbf{N\textsubscript{ch} (With CR)} & \textbf{N\textsubscript{ch} (Without CR)} \\
\hline
I     & 0--5    & 2346--3350 & 3254--4000 \\
II    & 5--10   & 1973--2346 & 2759--3254 \\
III   & 10--20  & 1397--1972 & 1935--2759 \\
IV    & 20--30  & 1009--1397  & 1341--1935 \\
V     & 30--40  & 674--1009   & 887--1341\\
VI    & 40--50  & 425--674   & 547--887 \\
VII   & 50--60  & 245--425   & 303--546 \\
VIII  & 60--70  & 108--224   & 144--303 \\
IX    & 70--90  & 14--117    & 15--144 \\
\hline

\end{tabular}
}
\end{table}
The ALICE data are available in terms of rapidity rather than pseudorapidity, and therefore, used the same variable in the model. To study the balance functions for different particle species, specific selection criteria are applied to match the kinematic ranges used in the experimental ALICE data. Pions are selected within the rapidity interval $-0.8 < y < 0.8$ and transverse momentum range $0.2 < p_{\mathrm T} < 2.0~\mathrm{GeV}/c$. Kaons are analyzed in the rapidity range $-0.7 < y < 0.7$ with the same transverse momentum interval, $0.2 < p_{\mathrm T} < 2.0~\mathrm{GeV}/c$. For protons, the rapidity window $-0.6 < y < 0.6$ is used, and the transverse momentum is $0.5 < p_{\mathrm T} < 2.5~\mathrm{GeV}/c$.

\section{Results and Discussion}

\begin{figure*}[ht!]
    \centering
    \includegraphics[width=0.9\textwidth]{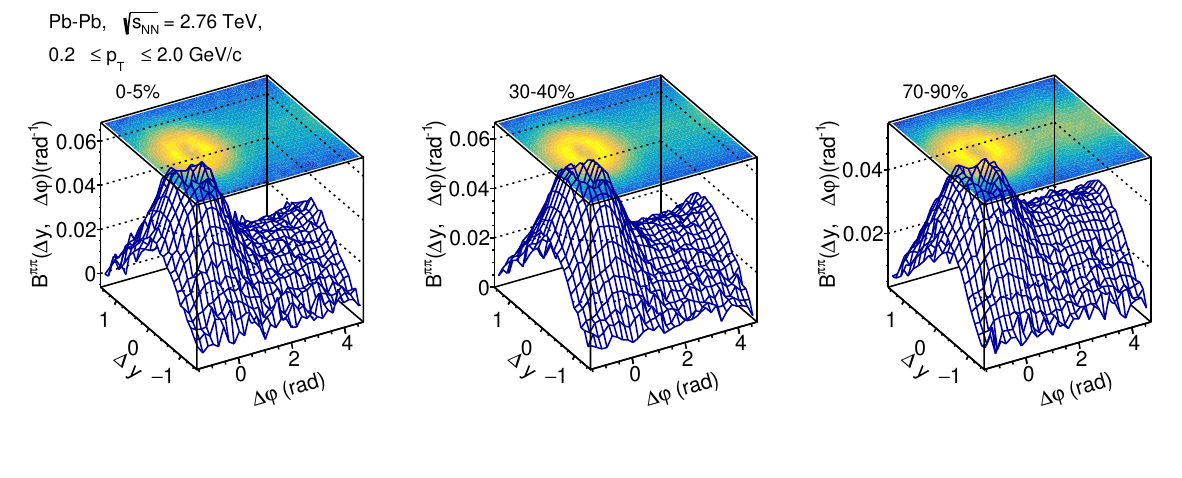}
    \caption{Balance functions $(B(\Delta y, \Delta \varphi))$ for pions in Pb--Pb collisions at $\sqrt{s_{NN}} = 2.76$ TeV for different centrality intervals: 0--5\%, 30--40\%, and 70--90\%.}
    \label{fig:balance}
\end{figure*}

 \begin{figure*}[ht!]
    \centering
    \includegraphics[width=0.8\textwidth]{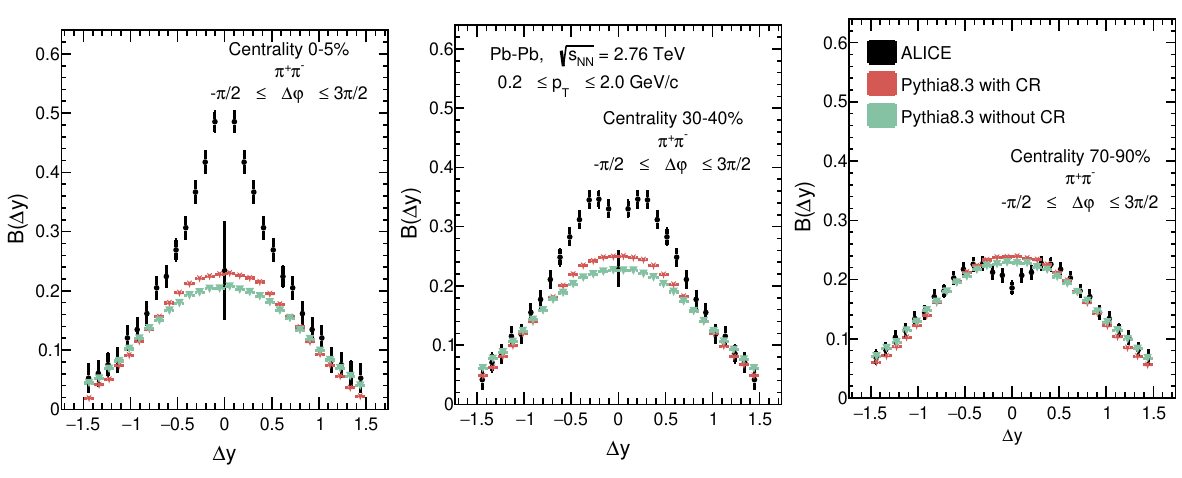}
    \caption{Balance functions $(B(\Delta y))$ for pions in Pb--Pb collisions at $\sqrt{s_{NN}} = 2.76$ TeV for different centrality intervals: 0--5\%, 30--40\%, and 70--90\% obtained using the Pythia8.3+ Angantyr model with and without color reconnection, and compared with the ALICE results.}
    \label{fig:balance}
\end{figure*}

 \begin{figure*}[ht!]
    \centering
    \includegraphics[width=0.8\textwidth]{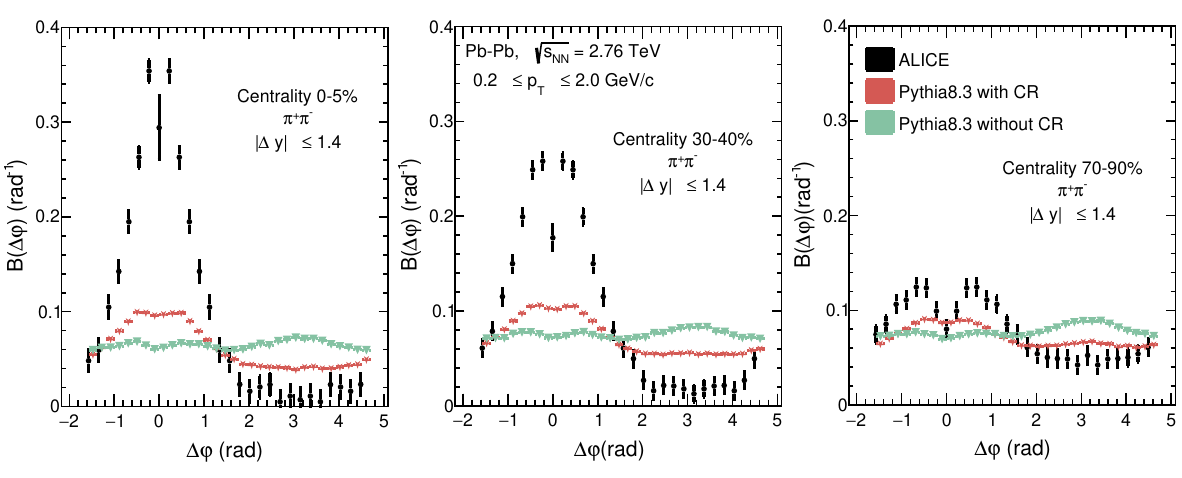}
    \caption{Balance functions $(B(\Delta \varphi))$ for pions in Pb--Pb collisions at $\sqrt{s_{NN}} = 2.76$ TeV for different centrality intervals: 0--5\%, 30--40\%, and 70--90\% obtained using the Pythia8.3 + Angantyr model with and without color reconnection, and compared with the ALICE results.}
    \label{fig:balance}
\end{figure*}

 \begin{figure*}[ht!]
    \centering
    \includegraphics[width=0.95\textwidth]{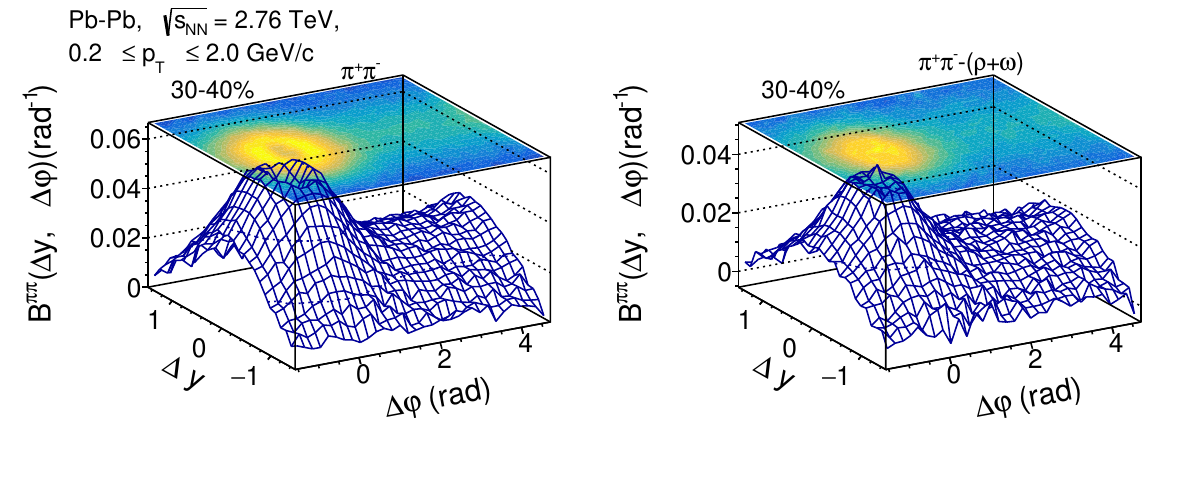}
    \caption{Balance functions \( B(\Delta y, \Delta\varphi) \) of pion pairs, for Pb--Pb collisions at $\sqrt{s_{NN}} = 2.76$ TeV in the  30--40 \% centrality class, obtained using the Pythia8.3 + Angantyr model with  color reconnection before (left) and after (right) the removal of $\rho$ and $\omega$ resonance.}
    \label{fig:balance}
\end{figure*}

 \begin{figure*}[ht!]
    \centering
    \includegraphics[width=0.95\textwidth]{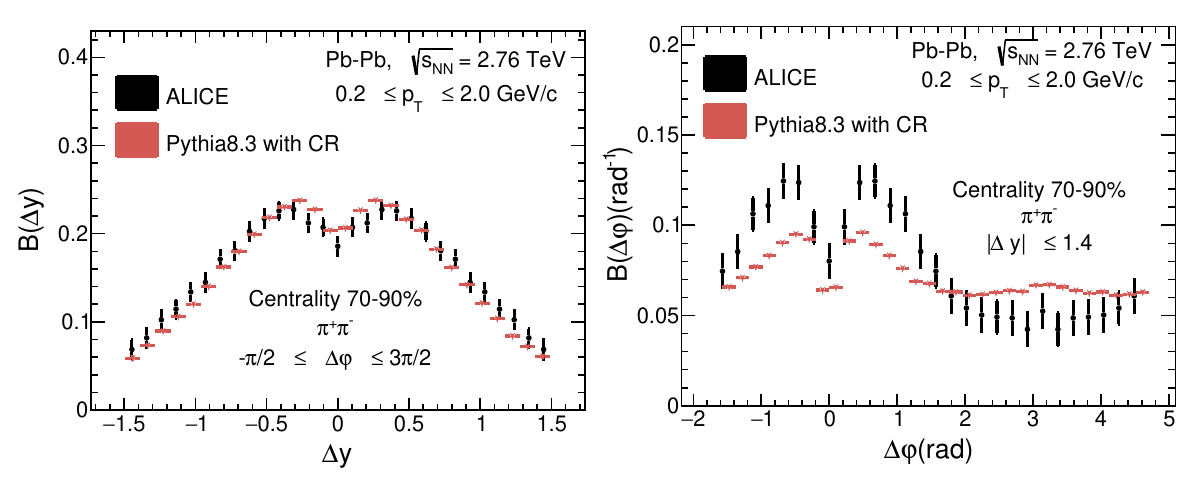}
    \caption{Balance functions of pion pairs, projected onto the  $\Delta y$ axis  (left) and the $\Delta \varphi$ axis (right), for Pb--Pb collisions at $\sqrt{s_{NN}} = 2.76$ TeV in the  70--90\% centrality class, obtained using the Pythia8.3 + Angantyr model with  color reconnection including Bose–Einstein correlations, and compared with the ALICE results.}
    \label{fig:balance}
\end{figure*}

\begin{figure*}[ht!]
    \centering
    \includegraphics[width=0.84\textwidth]{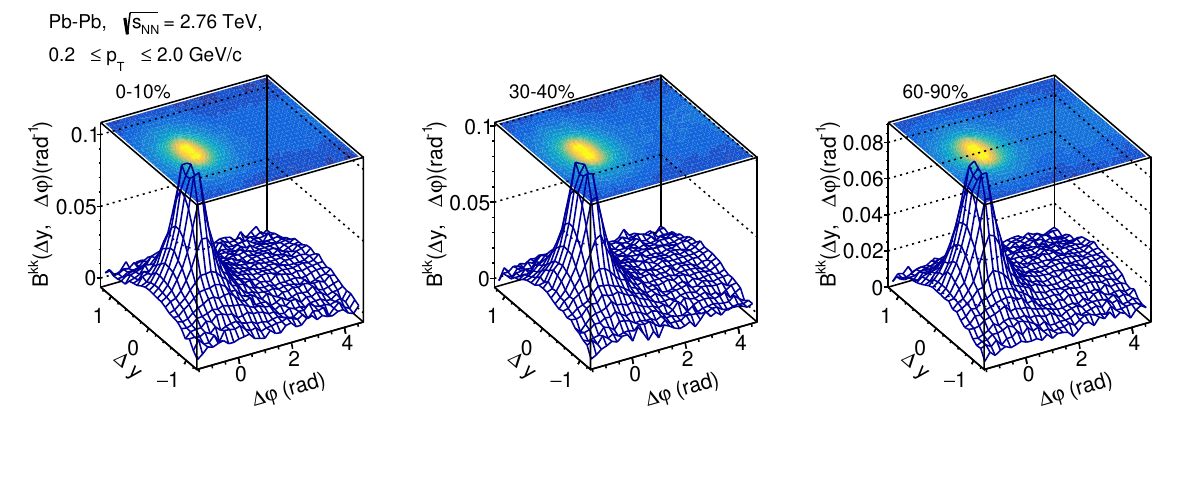}
    \caption{Balance functions $(B(\Delta y, \Delta \varphi))$ for kaons in Pb--Pb collisions at $\sqrt{s_{NN}} = 2.76$ TeV for different centrality intervals: 0--10\%, 30--40\%, and 60--90\%.}
    \label{fig:balance}
\end{figure*}

 \begin{figure*}[ht!]
    \centering
    \includegraphics[width=0.8\textwidth]{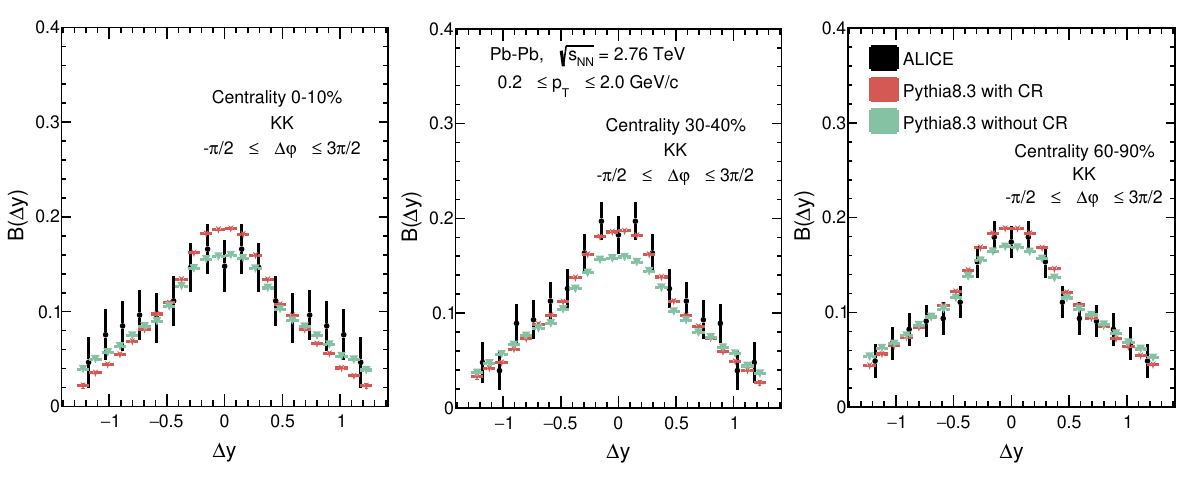}
    \caption{Balance functions $(B(\Delta y))$ for kaons in Pb--Pb collisions at $\sqrt{s_{NN}} = 2.76$ TeV for different centrality intervals: 0--10\%, 30--40\%, and 60--90\% obtained using the Pythia8.3 + Angantyr model with and without color reconnection, and compared with the ALICE results.}
    \label{fig:balance}
\end{figure*}
\begin{figure*}[ht!]
    \centering
    \includegraphics[width=0.8\textwidth]{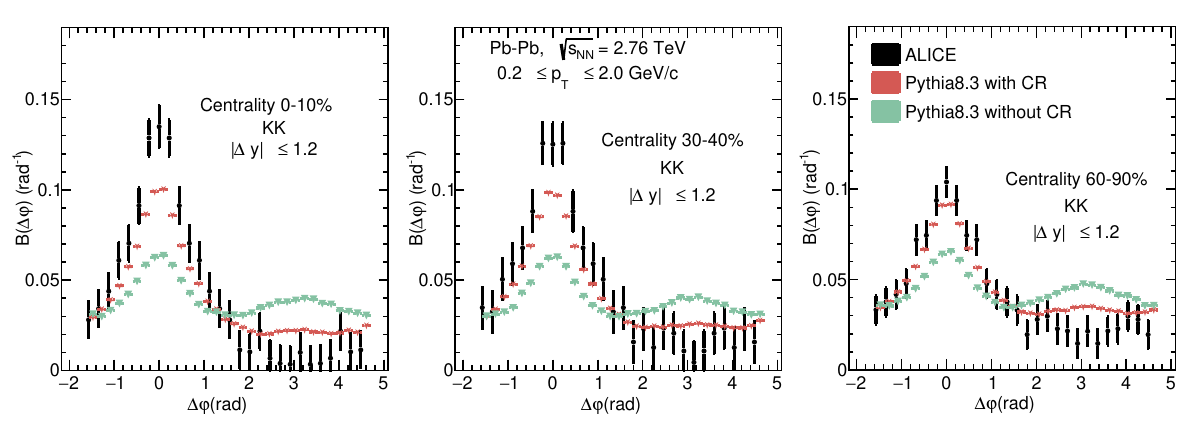}
    \caption{Balance functions $(B(\Delta \varphi))$ for kaons in Pb--Pb collisions at $\sqrt{s_{NN}} = 2.76$ TeV for different centrality intervals: 0--10\%, 30--40\%, and 60--90\% obtained using the Pythia8.3 + Angantyr model with and without color reconnection, and compared with the ALICE results.}
    \label{fig:balance}
\end{figure*}

\begin{figure*}[ht!]
    \centering
    \includegraphics[width=0.95\textwidth]{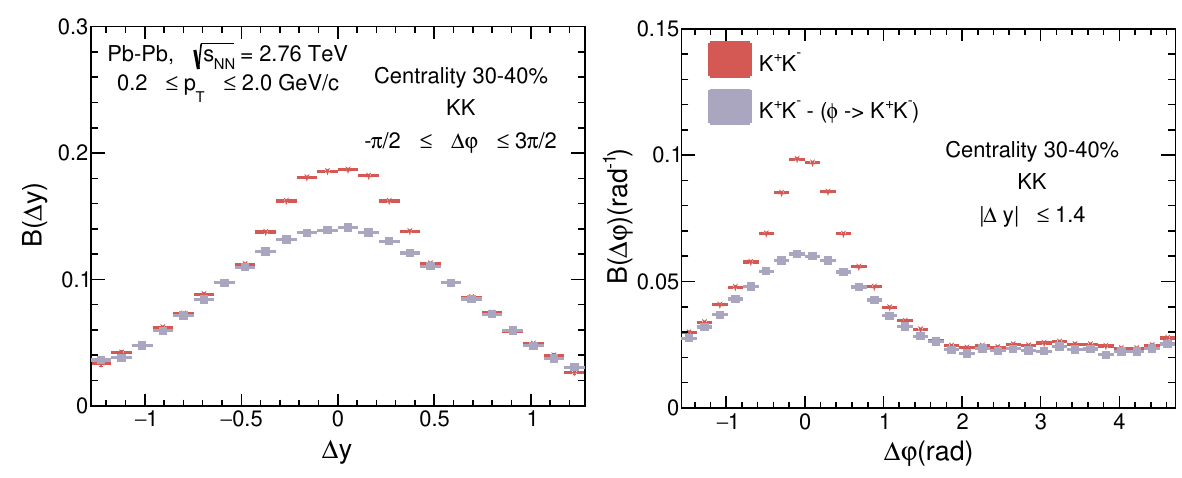}
      \caption{Balance functions of kaon pairs, projected onto the $\Delta y$ (left) and $\Delta\varphi$ (right) axes for Pb–Pb collisions at $\sqrt{s_{NN}} = 2.76$ TeV in the 30–40$\%$ centrality class, obtained using the Pythia8.3 + Angantyr model with color reconnection (CR), with and without resonance contributions.}
        \label{fig:balance}
\end{figure*}
\begin{figure*}[ht!]
    \centering
    \includegraphics[width=0.95\textwidth]{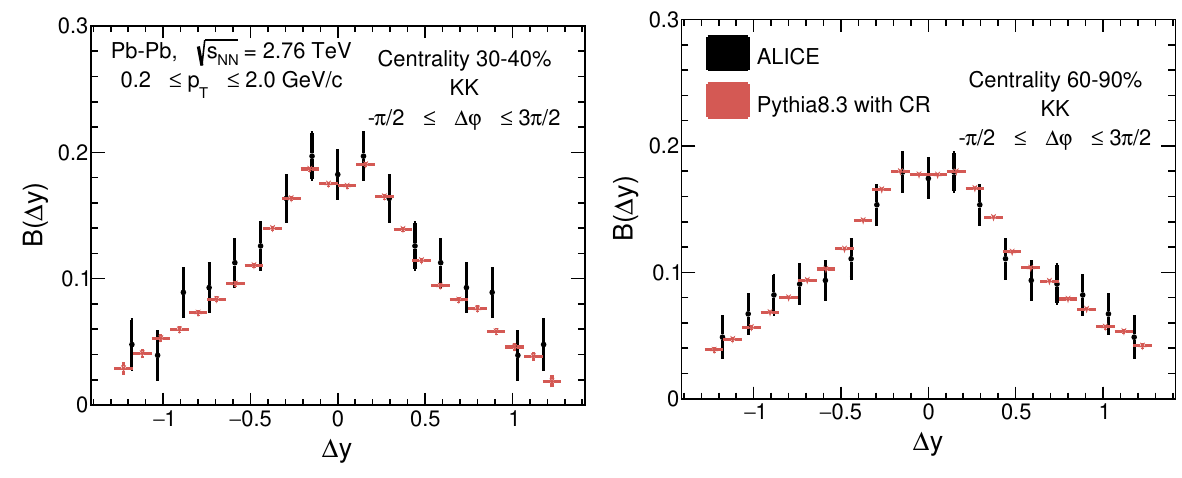}
    \caption{Balance functions of kaon pairs, projected onto the  $\Delta y$ axis for Pb--Pb collisions at $\sqrt{s_{NN}} = 2.76$ TeV in the  30--40\% (left) and 60--90\% (right) centrality classes, obtained using the Pythia8.3 + Angantyr model with  color reconnection including Bose–Einstein correlations, and compared with the ALICE results.}
    \label{fig:balance}
\end{figure*}

\begin{figure*}[ht!]
    \centering
    \includegraphics[width=0.85\textwidth]{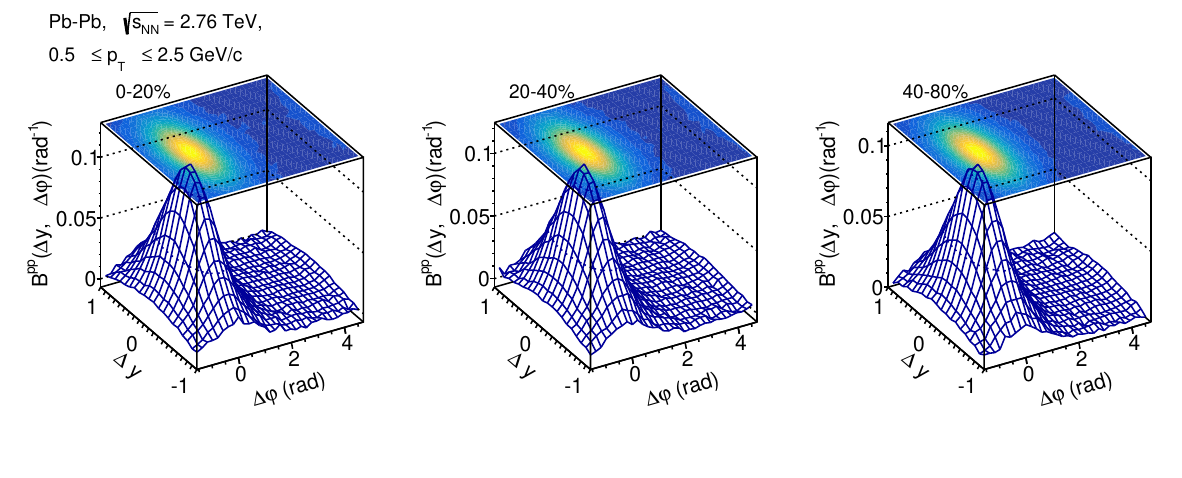}
    \caption{Balance functions $(B(\Delta y, \Delta \varphi))$ for protons in Pb--Pb collisions at $\sqrt{s_{NN}} = 2.76$ TeV for different centrality intervals: 0--20\%, 20--40\%, and 40--80\%.}
    \label{fig:balance}
\end{figure*}

\begin{figure*}[ht!]
    \centering
    \includegraphics[width=0.82\textwidth]{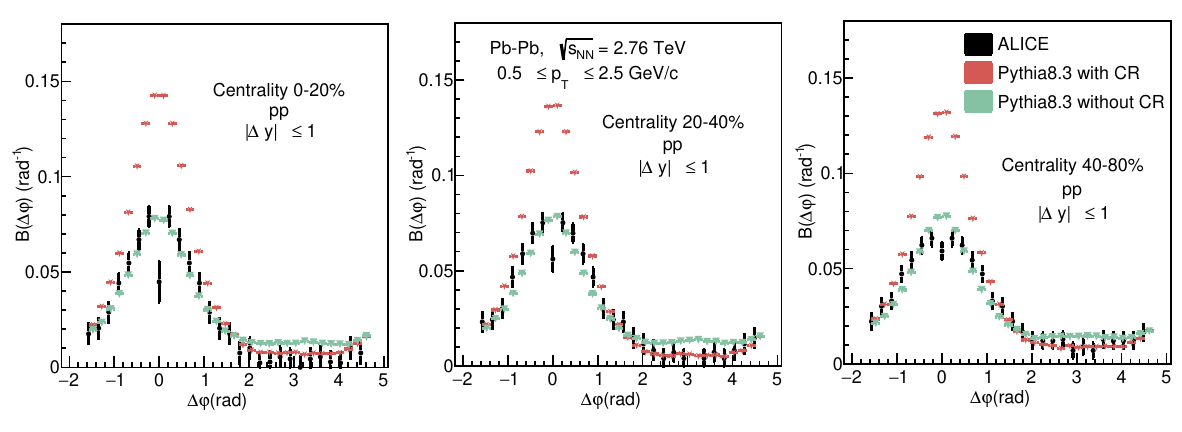}
    \caption{Balance functions $(B(\Delta \varphi))$ for protons in Pb--Pb collisions at $\sqrt{s_{NN}} = 2.76$ TeV for different centrality intervals: 0--20\%, 20--40\%, and 40--80\% obtained using the Pythia8.3 + Angantyr model with and without color reconnection, and compared with the ALICE results.}
    \label{fig:balance}
\end{figure*}

\begin{figure*}[ht!]
    \centering
    \includegraphics[width=0.82\textwidth]{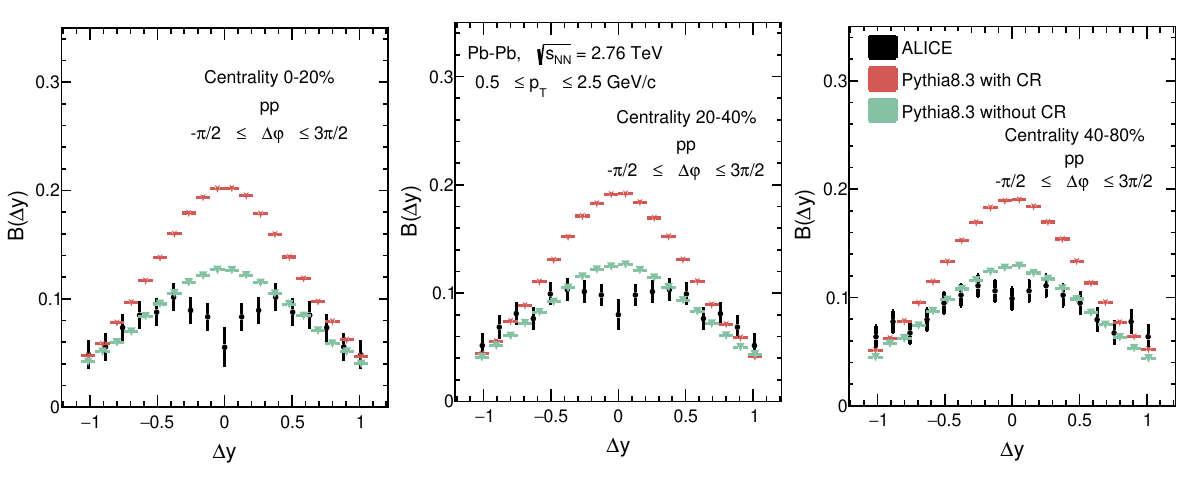}
    \caption{Balance functions $(B(\Delta y))$ for protons in Pb--Pb collisions at $\sqrt{s_{NN}} = 2.76$ TeV for different centrality intervals: 0--20\%, 20--40\%, and 40--80\% obtained using the Pythia8.3 + Angantyr model with and without color reconnection, and compared with the ALICE results.}
    \label{fig:balance}
\end{figure*}
Balance functions measures the correlation between the LS and US correlation. Figure 3 shows the BF 
B($\Delta y, \Delta \varphi$) for pions in Pb–Pb collisions, obtained using the Pythia8.3 + Angantyr model for different centrality classes: 0–5\%, 30–40\%, and 70–90\%. Figure 4 shows the BF 
B($\Delta y$) for pions in the same centrality intervals. It was obtained using the model with different tunes, both with and without CR, and compared with ALICE data. Similarly, figure 5 presents the BF 
B($\Delta \varphi$) for pions in these centrality ranges, calculated using the Monash 2013 tune within the Pythia + Angantyr framework and compared with ALICE data. As observed, for Pb–Pb collisions, the model, both with and without CR, does not quantitatively reproduce the experimental results for central collisions within the present implementation. In addition to the baseline configuration, we have also examined the impact of string shoving and rope hadronization within the PYTHIA8.3 + Angantyr framework. The inclusion of these mechanisms does not lead to any significant modification of the balance function shape or its centrality dependence within the current statistical precision of this study. However, the agreement is relatively better for peripheral collisions compared to central ones when CR is included. Without CR, the B($\Delta y$) shows a broader distribution with a smaller magnitude compared to the case with CR. Similarly, for B($\Delta \varphi$) the near side distribution is broader with a lower magnitude, and an additional correlation peak appears on the away-side, which is not observed in the experimental data. Therefore, the inclusion of CR provides a better description of the results, particularly for peripheral collisions.  As seen in figures 4 and 5 , the BF becomes wider when going from central to peripheral collisions and  B($\Delta y$) is better reproduced for peripheral collisions compared to  B($\Delta \varphi$). The observed discrepancies in central collisions may indicate the need for a dedicated heavy-ion tuning of the Angantyr framework.

In the experimental data \cite{Balance}, a dip is observed at $\Delta y$ = 0 and $\Delta \varphi$ = 0 which arises due to the effects of resonance decays and quantum statistical correlations among identical bosons. Now we discuss the effect of resonance decays on pions.  Pions originate from both resonance decays and non-resonance  particle production. A significant fraction of pions originate from the decay of resonances, mainly the $\rho$ and $\omega$ mesons, which decay into $\pi^{+}$ and $\pi^{-}$ pairs. These pions come from the same parent particle and contribute to the unlike-sign pairs in the BF. When these resonance contributions are removal, the dip structure observed at $\Delta y = 0$ and $\Delta \varphi = 0$ disappears completely and after the removal of resonance effect BF become narrower. Figure 6 shows the BF $B(\Delta y, \Delta \varphi)$ for pion pairs before (left) and after (right) the removal of resonance contributions for the 30--40\% centrality class. Thus, when the balance functions for pions is calculated using this  model before removing the resonance contributions, a small dip is observed at $\Delta y = 0$ and $\Delta \varphi = 0$.
   The Bose–Einstein correlation measures the correlation between identical bosons, and therefore, due to this effect, a dip appears  at  $\Delta y$ = 0 and $\Delta \varphi$ = 0.  When the BEC is explicitly included in Pythia, the dip structure can be described reasonably well for peripheral collisions but not for central collisions. This is because, in Pb–Pb collisions, the system size is significantly larger, while in the Pythia8.3 + Angantyr model the system size cannot be increased accordingly. Therefore, the model can reproduce the dip only for peripheral collisions, as shown in figure 7. The figure shows the $B(\Delta y)$(left) and $B(\Delta \varphi)$(right)  for Pb–Pb collisions, compared with the ALICE data. The dip observed in peripheral collisions is described by this model  with CR.      \\
   Since kaons contain strange quarks, and strange quarks are produced at the early stage of the collision, the BF for kaons is expected to be broader compared to that for pions. Figure 8 shows the BF ($B(\Delta y, \Delta \varphi)$) for kaon pairs in Pb–Pb collisions, obtained using this model  with CR for different centrality classes: 0–10\%, 30–40\%, and 60–90\%.   Figure 9 and figure 10 present the projection of the kaon BF on $\Delta y$ and $\Delta \varphi$ for different centrality use with CR and without CR. The results are compared with the ALICE data. It is observed that the kaon BF for Pb–Pb collisions can be described using this model with CR for all centrality classes since kaons are produced early and are therefore independent of centrality. Figure 11 presents the BF $B(\Delta y)$ (left) and $B(\Delta\varphi)$ (right) for kaons, obtained using the Pythia8.3 + Angantyr model with CR, both with and without resonance contributions. For kaons, the $B(\Delta y)$ and near-side $B(\Delta\varphi)$ are primarily dominated by contributions from $\phi$-meson decays, since $\phi$ meson predominantly decays into unlike-sign kaon pairs. When kaons originating from the same parent $\phi$ meson are excluded, the yield of $B(\Delta y)$ and the near-side $B(\Delta\varphi)$ decreases accordingly. Therefore, the kaon BF becomes wider after the removal of resonance contributions.
For kaons in the experimental data, a small dip is observed at $\Delta y = 0$. In contrast, the Pythia8.3 + Angantyr model without BEC does not reproduce this dip. However, when BEC are included, the model successfully describes the dip observed in the 30–40$\%$ and 60–90 $\%$ centrality classes. This can be attributed to the relatively smaller system size for kaons compared to pions. In central Pb–Pb collisions, the larger system size leads to a weaker manifestation of this effect, and the model fails to reproduce the dip for central collision. Figure 12 shows the projection of the kaon BF onto $\Delta y$ for the 30–40$\%$ (left) and 60–90$\%$ (right) centrality classes obtained using the Pythia8.3 + Angantyr model with color reconnection (CR) and Bose–Einstein correlations, compared with ALICE data. As shown, the inclusion of Bose–Einstein correlations improves the description of the dip at $\Delta y = 0$.  \\

\begin{figure*}[ht!]
    \centering
    \includegraphics[width=0.84\textwidth]{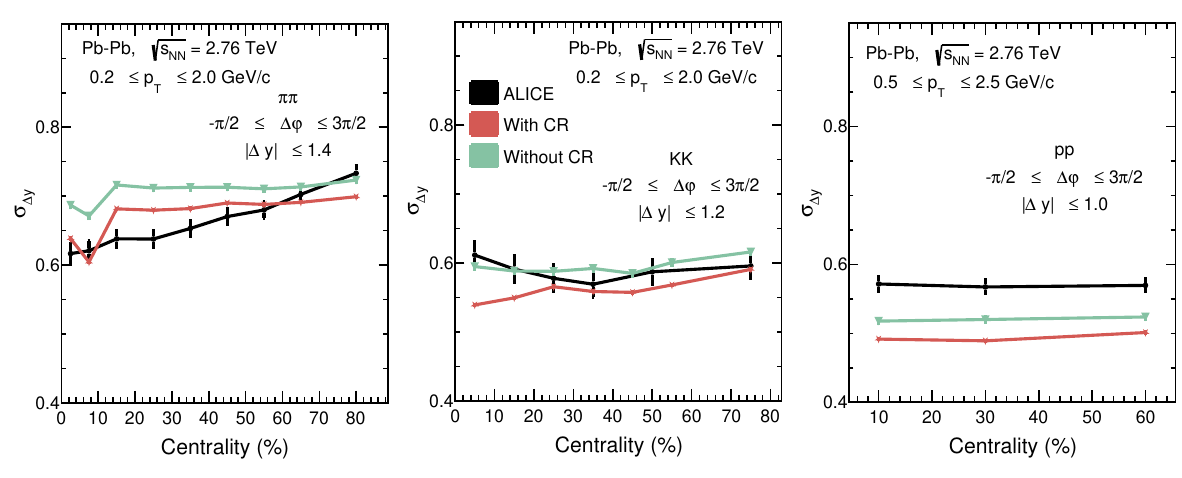}
    \caption{Balance functions width ($\sigma_{\Delta y}$) for pion pairs (left), kaon pairs (middle) and proton pairs (right) in Pb--Pb collisions at $\sqrt{s_{NN}} = 2.76$ TeV  used with Pythia8.3 + Angantyr and compared with ALICE data.}
    \label{fig:balance}
\end{figure*}

\begin{figure*}[ht!]
    \centering
    \includegraphics[width=0.84\textwidth]{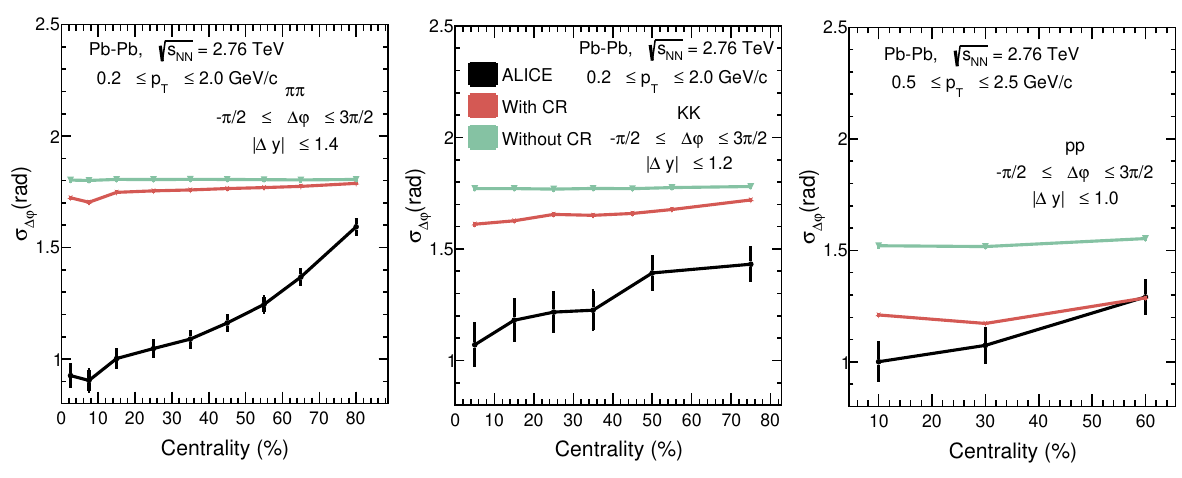}
    \caption{Balance functions width ($\sigma_{\Delta\varphi}$) for pion pairs (left), kaon pairs (middle) and proton pairs (right) in Pb--Pb collisions at $\sqrt{s_{NN}} = 2.76$ TeV  used with Pythia8.3 + Angantyr and compared with ALICE data.}
    \label{fig:balance}
\end{figure*}
\begin{figure*}[ht!]
    \centering
    \includegraphics[width=0.84\textwidth]{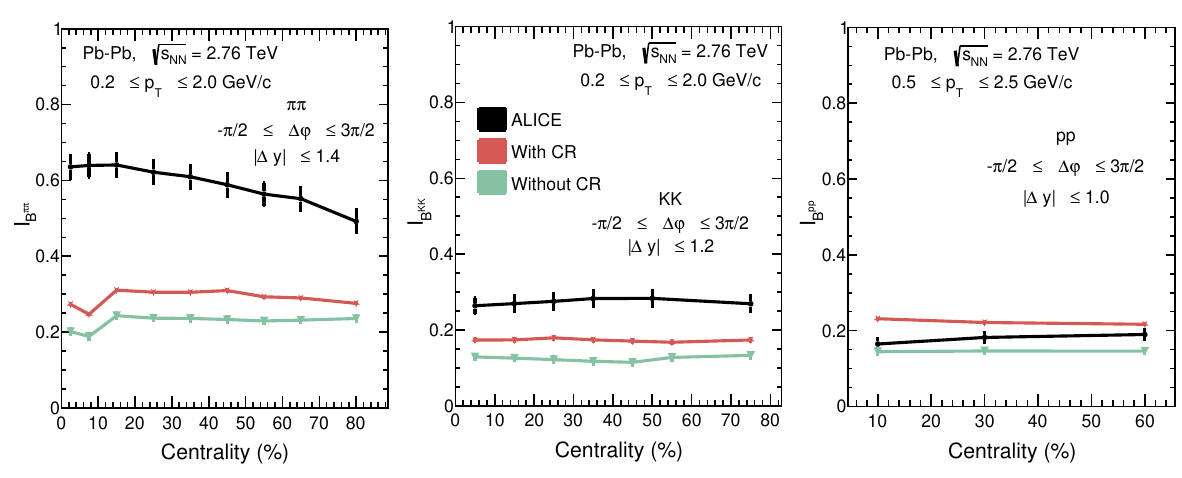}
    \caption{Balance functions Integral ($I_{B}$) for pion pairs (left), kaon pairs (middle) and proton pairs (right) in Pb--Pb collisions at $\sqrt{s_{NN}} = 2.76$ TeV  used with Pythia8.3 + Angantyr and compared with ALICE data.}
    \label{fig:balance}
\end{figure*}
Figure 13 shows the two-dimensional balance functions $B(\Delta y, \Delta\varphi)$ for protons in Pb–Pb collisions at different centrality classes: 0–20$\%$, 20–40$\%$, and 40–80$\%$, obtained using the Pythia8.3 + Angantyr model with CR. Figures 14 and 15 present the projections of the BF on the $\Delta\varphi$ and $\Delta y$ axes, respectively, both with and without CR, and are compared with the ALICE data. As shown in the figures, the results with CR do not reproduce the experimental data, whereas the results without CR provide a better description. A dip is observed in the ALICE data at $\Delta y = 0$ and $\Delta\varphi = 0$, This feature has been attributed to baryon–antibaryon interactions in the hadronic medium, including proton–antiproton annihilation as well as baryon regeneration processes ~\cite{protondip}. However, this  model does not include any hadronic rescattering or medium effects; therefore, this dip is not reproduced by the model.
The widths ($\sigma$)  of the BF for pion, kaon, and proton pairs in $\Delta y$ and $\Delta\varphi$ are shown in figures 16 and 17 for Pb–Pb collisions, obtained using the model and compared with the ALICE data. The widths are quantified using the RMS values. As seen from the experimental results, the pion BF shows a dependence on centrality, whereas the kaon and proton BF do not show any dependence on centrality. This behavior can be explained by the fact that kaons contain strange quarks, which are produced at the early stage of the collision, making their widths less sensitive to centrality. Similarly, the proton BF shows little to no centrality dependence, suggesting that baryon–antibaryon pairs are produced predominantly at an early stage. When described by the model, the pion BF exhibits a weak centrality dependence when CR is included, while without CR it shows almost no dependence on centrality. For kaons, $\sigma_{\Delta Y}$ centrality is consistent with the experimental uncertainties for with CR and beyond 20 $\%$ centrality without CR. The trends observed in $\sigma_{\Delta Y}$ and $\sigma_{\Delta \varphi}$ for kaons are reproduced by the model when CR is included. For kaons and protons, the widths remain nearly independent of centrality both with and without CR. A similar trend is observed for the integrated BF, which becomes narrower when moving from central to peripheral collisions for pion and independent on centrality , as shown in figure 18 for pions (left), kaons (middle), and protons (right). For pion integral BF using the with CR model small dependency on centrality, the Pythia8.3 + Angantyr model does not fully reproduce the measured widths and integrals of the BF for Pb–Pb collisions.
\section{Summary}
\label{sum}
In this work, we study the balance functions for identical particles in Pb–Pb collisions at $\sqrt{s_{NN}} = 2.76$ TeV using the Pythia8.3 + Angantyr event generator. This study focuses on investigating the contributions of resonances and Bose–Einstein correlations for bosons. The main observations of this work are summarized below:

\begin{itemize}
\setlength{\itemsep}{2.0pt}
\setlength{\parskip}{2.0pt}
\setlength{\parsep}{2.0pt}
    \item  
    The BF for pions become narrower when moving from peripheral to central Pb–Pb collisions. Using the Pythia8.3 event generator coupled with the Angantyr model and color reconnection, the results show good agreement with ALICE data for peripheral collisions; however, for the Monash 2013 tune employed in this study, the model does not quantitatively reproduce the BF in central collisions. This indicates that the present implementation, without a dedicated heavy-ion tuning of the Angantyr framework, may be insufficient to fully describe central Pb–Pb data.  For the pion BF, a small dip is observed at $\Delta y = 0$ and $\Delta \varphi = 0$ when using the  model. This dip arises due to the effect of resonance decays, and it disappears when the resonance contribution is removed. In the experimental data, a dip is also observed at $\Delta y = 0$ and $\Delta \varphi = 0$. This feature can be described using the model when both Bose–Einstein correlations and resonance effects are included, however it is only  for peripheral collisions, not for central collisions.
\item For kaons, the BF remains nearly unchanged when going from peripheral to central collisions. Using the Pythia8.3 + Angantyr event generator, the results show good agreement with the ALICE data when CR is included. For kaons, the resonance contribution is larger compared to other particle species, which dominates the  B($\Delta y$) and  near side B($\Delta \varphi$) distributions. The small dip observed in B($\Delta y$) at 
$\Delta y =0 $ is well explained by the  model when the BEC is included, particularly for the 30--40$\%$ and 60--90$\%$ centrality classes.
\item For protons, the BF is independent of centrality from peripheral to central collisions. Using the Pythia8.3 + Angantyr event generator reconnection shows that baryon–antibaryon pairs are produced at an early stage of the collision. The results show good agreement with the ALICE data when color reconnection is turned off.

\item The balance functions widths in $\Delta y$ and $\Delta \varphi$, as well as the integrals obtained using the  event generator, do not fully describe the experimental data. However, the model shows that for pions, the width does not depend on centrality without CR, while a small centrality dependence appears when CR is included, with the width becoming broader from central to peripheral collisions and integral BF decrease  from  central to peripheral collisions for pion.  For kaons and protons, the BF widths  and integral BF show no dependence on centrality, which consistent with the experimental observations.

\end{itemize}

\section*{Acknowledgement} 
R.G. acknowledges the financial support provided by the Council of Scientific
and Industrial Research (CSIR) (File
No.09/1022(13483)/2022-EMR-I).

\bibliographystyle{IEEEtran}
\nocite{*}
\bibliography{refs}

 \end{document}